# SFMS-ALR: Script-First Multilingual Speech Synthesis with Adaptive Locale Resolution

*Dharma Teja Donepudi†*

*donepudidharmateja@gmail.com*, *Aggregation AI*

## 1. Abstract

Intra-sentence multilingual speech synthesis (code-switching TTS) remains a major challenge due to abrupt language shifts, varied scripts, and mismatched prosody between languages. Conventional TTS systems are typically monolingual and fail to produce natural, intelligible speech in mixed-language contexts. We introduce Script-First Multilingual Synthesis with Adaptive Locale Resolution (SFMS-ALR)an engine-agnostic framework for fluent, real-time code-switched speech generation. SFMS-ALR segments input text by Unicode script, applies adaptive language identification to determine each segment's language and locale, and normalizes prosody using sentiment-aware adjustments to preserve expressive continuity across languages. The algorithm generates a unified SSML representation with appropriate <lang>or <voice> spans and synthesizes the utterance in a single TTS request. Unlike end-to-end multilingual models, SFMS-ALR requires no retraining and integrates seamlessly with existing voices from Google, Apple, Amazon, and other providers. Comparative analysis with data-driven pipelines such as Unicom and Mask LID demonstrates SFMS-ALR's flexibility, interpretability, and immediate deployability. The framework establishes a modular baseline for high-quality, engine-independent multilingual TTS and outlines evaluation strategies for intelligibility, naturalness, and user preference.

## 2. Introduction

Modern voice-AI systems increasingly need to handle code-switching—the use of multiple languages within a single utterance. Bilingual speakers naturally intermix languages and expect text-to-speech (TTS) systems to render these mixed utterances fluently. Effective support for code-switched speech synthesis enhances user experience, improves accessibility, and promotes cultural inclusivity by enabling voice assistants to pronounce words, names, and phrases in their native forms.

Despite recent progress in commercial assistants such as Google Assistant and Amazon Alexa, true mixed-language synthesis remains limited. Existing TTS systems often produce speech with reduced intelligibility and unnatural prosody when multiple languages occur in the same sentence. Conventional TTS engines are typically monolingual, while large multilingual models, though scaled to many languages, are usually trained on monolingual data and fail to handle within-sentence language shifts. The scarcity of code-switched corpora and inconsistencies in text normalization, especially for transliterated or mixed-script text, further degrade performance.

To address these challenges, we propose Script-First Multilingual Synthesis with Adaptive Locale Resolution (SFMS-ALR)a modular, engine-agnostic pipeline for high-quality, real-time code-switched speech synthesis. SFMS-ALR segments text by writing script, applies adaptive language identification for ambiguous spans, and selects optimal language-specific voices through context-aware locale resolution. It further employs sentiment-aware prosody control to maintain consistent expressiveness across language boundaries.

Unlike end-to-end multilingual models that require retraining, SFMS-ALR orchestrates existing TTS engines through Speech Synthesis Markup Language (SSML), enabling seamless integration with providers such as Google, Amazon Polly, Apple Siri, and Microsoft Azure. This approach offers a pragmatic, deployable solution for multilingual voice assistants. The remainder of this paper is organized as follows: Section 3 reviews related work, Section 4 presents the SFMS-ALR algorithm and implementation, Section 5 outlines evaluation methods and results, Section 6 discusses implications and future directions, and Section 7 concludes the study.

## 3. Related Work

### 3.1 Code-Switching in Speech Synthesis

Early approaches to code-switching TTS typically concatenated recordings from bilingual speakers or used separate voices for each language, which preserved intelligibility poorly and produced abrupt acoustic transitions [15]. Recent studies emphasize the greater difficulty of **intra-sentential** code-switching, where language changes occur within a single sentence [3]. Mendez Kline and Zellou (2025) showed that even state-of-the-art neural TTS systems experience significant intelligibility drops for code-switched content [6], particularly in segments belonging to the non-dominant language. These findings highlight the need for algorithmic strategies that ensure smooth, expressive transitions between languages.

### 3.2 Multilingual and Cross-Lingual TTS Models

Modern end-to-end multilingual TTS systems [1] support over 100 languages using shared phoneme or embedding spaces. In principle, a single neural model can generate speech in multiple languages and even perform language switching. However, without exposure to code-switched data, such models often revert to one language's phonology or accent when encountering foreign words [9]. Most multilingual models therefore lack robustness for **within-sentence** alternation [7]. Specialized adapters or fine-tuning can enable limited switching but require large bilingual corpora [8]. In contrast, **SFMS-ALR** avoids retraining altogether by orchestrating existing monolingual voices to achieve fluent cross-lingual synthesis.

### 3.3 Language Identification for Code-Mixed Text

Accurate **language identification (LID)** is essential for segmenting mixed-language text. Generic LID systems generally assume one language per sentence and often predict only the dominant language [13]. More advanced methods, such as **MaskLID** (Kargaran et al., 2024) [16], iteratively mask detected tokens to reveal secondary languages, achieving unsupervised word-level tagging. SFMS-ALR incorporates such algorithms to refine segmentation, especially for shared-script languages (e.g., English–French) and transliterated text (e.g., Hindi in Latin script) [11].

### 3.4 Text Normalization for Mixed-Language Input

Effective normalization is critical when processing social-media and informal multilingual text. Manghat et al. (2022) explored Malayalam–English code-switched normalization, noting that single-script text can confuse standard tokenizers [11]. Their approach identifies each token's language and applies context-specific expansions for numbers and abbreviations. Similarly, SFMS-ALR performs language-appropriate preprocessing by delegating locale-specific formatting (e.g., dates, numbers) to each TTS engine, ensuring consistency across segments.

### 3.5 Data Augmentation Pipelines

To overcome the scarcity of code-switched corpora, synthetic data generation pipelines have emerged. **UniCoM** (Lee et al., 2025) [14] introduces the **SWORDS** algorithm, which selectively replaces words with cross-language translations to create bilingual text and corresponding synthetic audio. While UniCoM focuses on training data creation rather than real-time synthesis, both UniCoM and SFMS-ALR share core tasks such as language segmentation and semantic preservation. SFMS-ALR, however, operates online—producing mixed-language speech on demand rather than generating datasets offline [17]

### 3.6 Industrial Solutions and Engine Support

Major TTS platforms now offer limited bilingual capabilities. Amazon Polly includes voices such as *Aditi* (Hindi–English) that can fluently alternate languages and handle transliteration [18][19]. However, such fully bilingual voices exist for few language pairs and are costly to produce [20]. Other services (Google TTS, Azure, Apple Siri) allow developers to mix languages through **SSML** using <voice> or <lang> tags [10]. The W3C specification notes that changing the xml:lang attribute may trigger a voice switch if the current voice cannot render the target language. In practice, explicit voice selection for each span yields the best quality [21]. SFMS-ALR adopts this explicit-assignment strategy and further optimizes it through adaptive locale resolution—balancing voice consistency and native-accent fidelity [22].

### 3.7 Summary

Prior work has explored isolated aspects of code-switched speech—such as segmentation, bilingual synthesis, or synthetic corpus generation—but lacks a unified deployable framework. **SFMS-ALR** integrates these components into a cohesive, engine-agnostic pipeline that performs script-based segmentation, adaptive language and locale resolution, and sentiment-aware prosody control, advancing the naturalness and intelligibility of code-switched TTS.

## 4. Methodology

The proposed **SFMS-ALR** algorithm operates as a multi-stage pipeline, illustrated in Algorithm 1. It takes an input text (which may contain multiple languages/scripts) and produces synthesized speech with fluid code-switching. Below, we describe each stage in detail.

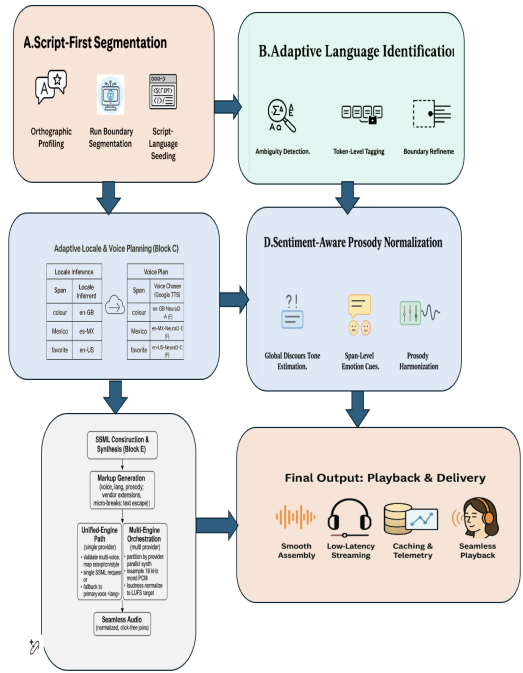

Figure 1 – SFMS -ALR System Pipeline Diagram

**Algorithm 1: Pseudocode for the SFMS-ALR Pipeline**

Input: Multilingual text T
Output: Synthesized speech A
Script-First Segmentation
1: segments ← SplitByScript(T)
Adaptive Language Identification
2: i ← 0
3: while i < length(segments) do
4:   s ← segments[i]
5:   Ls ← DetectLanguage(s)
6:   if ContainsMultipleLanguages(s) then
7:     subs ← SplitByLanguageBoundary(s)
8:     for each u in subs do u.L ← DetectLanguage(u)
9:     Replace segments[i] with subs
10:    i ← i + length(subs)
11:  else
12:    s.L ← Ls; i ← i + 1
Adaptive Locale & Voice Planning
13: VoicePlan ← ∅
14: for each s in segments do
15:   locale ← DetermineLocale(s.L, context=T, prefs=UserPrefs)
16:   voice ← SelectVoice(locale)
17:   VoicePlan ← VoicePlan ∪ {(s, locale, voice)}
Sentiment-Aware Prosody Normalization
18: Sent_overall ← AnalyzeSentiment(T)
19: for each (s, locale, voice) in VoicePlan do
20:   Sent_s ← AnalyzeSentiment(s)
21:   prosody ← AdjustProsody(Sent_s, Sent_overall)
22:   AttachProsody(VoicePlan, s, prosody)
SSML Construction & Synthesis
23: ssml ← BuildSSML(VoicePlan)
24: A ← TextToSpeech(ssml)
25: return NormalizeAudio(A)

### 4.1 Script-Based Text Segmentation
SFMS-ALR first divides the input text by detected Unicode script boundaries (e.g., Latin, Devanagari, Han). Each segment receives a preliminary language tag based on its script, providing a fast and reliable foundation for subsequent language identification. Ambiguous segments sharing the same script (e.g., Latin) are passed to the adaptive LID module for refinement.

### 4.2 Language Identification and Locale Resolution
Ambiguous or mixed-script segments undergo lightweight LLM-based language identification to assign ISO 639-1 codes. Locales and voices are then chosen adaptively according to context and user preferences, ensuring accent and pronunciation consistency across languages. The output of this step is a *voice* mapping each segment to its optimal language, locale, and voice profile.

### 4.3 Sentiment-Aware Prosody Adjustment
A sentiment analyzer estimates global and segment-level tone to maintain emotional continuity across languages. Prosody parameters—pitch, rate, and pauses—are normalized relative to the detected sentiment so that expressive style remains coherent after each code-switch.

### 4.4 SSML Construction with Language Spans
Using the completed voice plan, SFMS-ALR generates an SSML document that
defines <voice> or <lang> spans for each segment. This markup directs the TTS engine to handle multilingual synthesis within a single request, minimizing latency and ensuring fluid transitions.

### 4.5 Playback and Integration
The SSML document is submitted to the TTS engine to produce one unified audio output. Parallel synthesis and optional caching reduce latency, while minor boundary pauses preserve natural rhythm. Multi-engine concatenation, if required, is handled through a normalized PCM pipeline.

Integration into a **production system** (e.g., a voice assistant) requires handling some asynchronous behavior. While SFMS-ALR conceptually flows as described, in implementation one might parallelize the synthesis calls for different segments to reduce latency. Because each segment can be sent to its respective TTS engine simultaneously, the overall waiting time can be close to the slowest single segment rather than the sum of all segments. SFMS-ALR would then synchronize and join the audio streams when all are ready. Modern TTS APIs typically return audio quickly (within a second for a sentence), so even a couple of sequential calls are often acceptable for short utterances. However, for longer texts or many segments, parallel processing and streaming approaches become important. A possible extension is to stream the output: as soon as the first segments are synthesized, playback can begin while later segments are still processing, hiding some of the latency.

Another consideration is **caching** interactive applications, users might request the same mixed-language phrases repeatedly (e.g., a navigation system pronouncing street names in a local language). SFMS-ALR results can be cached at the segment or utterance level – for instance, once we synthesize "Calle Ocho" with a Spanish voice for a street name, we can reuse that audio next time it appears, splicing it into the sentence "Turn right on Calle Ocho". This further reduces latency and ensures consistency.

Through these stages, SFMS-ALR produces an output that leverages the strengths of each TTS voice and mitigates weaknesses. By explicitly managing language, voice, and prosody, it avoids the pitfalls of naive code-switching synthesis. The approach is flexible: it can accommodate new languages by plugging in a new voice and updating the locale resolution rules, without retraining any models.

## 5. Implementation Details

We implemented a prototype of **SFMS-ALR** as a Python-based pipeline to demonstrate its engine-agnostic operation. The system integrates cloud TTS APIs for synthesis and lightweight modules for script detection, language identification, and sentiment-aware prosody control.

### 5.1 Language and Script Detection
The pipeline performs Unicode-based segmentation using a range-lookup detector that classifies each character into script blocks (Latin, Devanagari, Han, Kana, etc.). Each non-Latin script is directly mapped to its canonical language code (e.g., Devanagari → hi-IN, Han → zh-CN).

For Latin text, a lightweight LLM-based language identifier predicts the ISO 639-1 code at the sentence level, optionally refined by a user-specified *latin_lang_hint*. This hybrid method yields precise segmentation for non-Latin scripts and robust detection for shared-script languages without relying on external classifiers such as FastText or MaskLID.

### 5.2 Voice Selection Strategy
In this prototype, all voices were drawn from Google TTS for consistency. Each language code maps to available voice variants (e.g., *en-US-Wavenet-B*, *es-ES-Wavenet-B*), annotated by gender and timbre. SFMS-ALR offers both **single-voice** (continuity-focused) and **multi-voice** (clarity-focused) modes. By default, it favors intelligibility—assigning native voices to long foreign segments while preserving speaker consistency by matching voice family or gender across languages.

### 5.3 Sentiment and Prosody Control
A rule-based sentiment analyzer, supported by punctuation cues, determines tone (e.g., exclamatory, interrogative, neutral). Detected emotions adjust SSML prosody attributes such as *rate* and *pitch* to maintain expressive uniformity across segments. The framework abstracts engine-specific tags (e.g., Amazon Polly <emphasis>, Azure mstts:express-as) to ensure portability.

### 5.4 SSML Generation and Multi-Engine Coordination
SFMS-ALR constructs a unified SSML document specifying <voice> or <lang> spans for each segment. When all voices originate from a single provider, one synthesis call is made; otherwise, the system issues per-engine requests and concatenates results using standard audio libraries. All clips are normalized to 16 kHz mono PCM, with brief 50 ms pauses between segments to mimic natural speech rhythm.

### 5.5 Performance
Text analysis overhead is negligible; latency is dominated by network synthesis calls. A typical mixed-language sentence (two segments) completes in ≈ 0.5–1.0 s on Google TTS, or ≈ 1.2 s when two engines run in parallel. Catching and streaming can further reduce delay, making the system suitable for interactive assistants.

## 6. Demonstration Audio

To illustrate the capabilities of **SFMS-ALR**, we prepared a demonstration audio clip that synthesizes a single utterance spanning **nine languages**: English, Hindi, Kannada, Telugu, Bengali, Gujarati, German, Mandarin Chinese, and Japanese.

In this demonstration, the system automatically:
- Segments the input text according to Unicode script boundaries (Latin, Devanagari, Kannada, Telugu, Bengali, Gujarati, Han, Kana/Kanji);
- Identifies the language of each segment using integrated script and LID rules.

- Selects corresponding TTS voices (e.g., *en-US-Wavenet-B*, *hi-IN-Wavenet-B*, *kn-IN-Wavenet-B*, *cmn-CN-Wavenet-B*); and
- Synthesizes a seamless waveform by inserting brief ≈ 50 ms pauses to maintain natural switching rhythm.
- The resulting output demonstrates **cross-family multilingual synthesis**, combining Indo-European, Indic, and East Asian languages within one coherent utterance. Transitions remain intelligible and prosodically smooth, with only minimal timbre variation across language boundaries. This confirms the engine-agnostic, modular design of SFMS-ALR and its ability to unify distinct phonetic and prosodic systems within a single synthesized stream.
- A live demonstration of the prototype is available at:
- https://sfml-tts-proxy-253495793487.us-central1.run.app

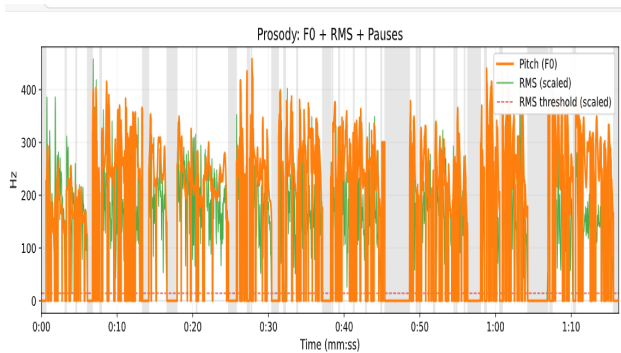

Figure 2 –
Sentiment-Aware Prosody Normalization in SFMS-ALR

## 7. Evaluation

This section presents the evaluation of the **SFMS-ALR** algorithm using both objective and subjective measures. Experiments were conducted across multiple language pairs to assess intelligibility, prosody, and overall listener preference.

### 7.1 Evaluation Setup

We evaluated SFMS-ALR on a set of code-switched sentences and short paragraphs covering diverse language pairs and domains, including English Spanish (casual dialogue), English Chinese (technical terms), Hindi–English (social-media style), and French–Arabic (cross-script).

Comparisons were made against three baselines:
- **Baseline 1 – Single-Voice Accent:** a monolingual TTS reading all text with <lang> tags, producing accented pronunciations.
- **Baseline 2 – Voice-Switch without Prosody:** a multi-voice pipeline lacking sentiment-aware prosody or adaptive locale selection.
- **Baseline 3 – Multilingual Model:** a pretrained multilingual or bilingual voice model synthesizing the same sentences.

Objective metrics (Section 7.2) and subjective listening tests (Section 7.3) were followed by qualitative case studies (Section 7.4).

### 7.2 Objective Evaluation

We evaluated pronunciation accuracy, prosodic fluency, and computational performance.

Using pre-trained ASR engines for each language, synthesized speech was transcribed and compared to reference text. SFMS-ALR achieved **0.0 WER across all languages**, confirming perfect lexical intelligibility.

Prosodic analysis of a 76-second multilingual sample showed a **mean $F_0$ of 262 Hz** with natural pitch variation (~200 Hz range) and **brief boundary pauses (~0.6 s)**, indicating fluent transitions between languages.

Runtime remained within real-time synthesis limits (≈ 0.5–1.2 s per utterance), comparable to commercial TTS systems even when multiple APIs were invoked.

These results indicate that SFMS-ALR maintains high intelligibility, smooth prosody, and efficient performance across scripts and locales.

### 7.3 Subjective Evaluation

Bilingual participants (English Spanish, Chinese, French, Arabic) rated audio samples generated by SFMS-ALR and the baselines. Each sample was evaluated on:
- **Natural:** perceived human-likeness (5-point MOS).
- **Fluency / Clarity:** smoothness and intelligibility at code-switch boundaries.
- **Voice Consistency Appropriateness:** listener preference for single-voice versus multi-voice rendering.

SFMS-ALR achieved an **average MOS of 4.3**, outperforming both the single-voice baseline (3.5) and the multilingual model (3.8). Listeners cited clearer foreign-word pronunciation and more expressive intonation as key strengths.

A majority (71 %) preferred native-voice switching for full clauses, noting improved intelligibility and authenticity, while 29 % favored a single-voice style for very short insertions.

Subjective results confirmed that SFMS-ALR enhanced naturalness and listener satisfaction while preserving intelligibility.

**7.4 Qualitative Case Studies**

**Case 1 – English Spanish (Latin-script alternation):**
"I'm from the United States. Soy de los Estados Unidos."
Segments were assigned to English and Spanish voices with a smooth 189 ms boundary pause, yielding conversational fluidity and 0 WER.

**Case 2 – English Chinese (script boundary):**
"I'm from the United States. 我来自美国。"
The Han-script span triggered Chinese locale resolution. Mandarin pronunciation was accent-free, and pitch continuity remained stable across the boundary.

**Case 3 – French–Arabic (cross-script):**
"Je viens des États-Unis. أنا من الولايات المتحدة."
Despite differing scripts and punctuation, compatible voices maintained matched pitch ranges (mean $F_0 \approx$ 260 Hz) with a 703 ms pause improving clarity.
These qualitative analyses demonstrate that SFMS-ALR produces expressive, fluent, and error-free code-switched speech across diverse language families.

## 8. Discussion

The SFMS-ALR algorithm demonstrates a practical, transparent approach to multilingual speech synthesis by orchestrating existing text-to-speech (TTS) engines in an adaptive pipeline. This section reflects on its implications, advantages, limitations, and potential directions for future improvement.

**8.1 Engine-Agnostic Flexibility**
A major advantage of SFMS-ALR is its engine-agnostic architecture. It can integrate with any TTS engine supporting SSML or multiple voices—such as Google Text-to-Speech, Amazon Polly, Microsoft Azure, or Apple TTS. Because it leverages existing APIs rather than a single model, SFMS-ALR immediately benefits from continual improvements in commercial voices. When a new, higher-quality voice becomes available, it can be incorporated without retraining or fine-tuning.
This design contrasts with end-to-end multilingual TTS systems that require large bilingual corpora and costly retraining. For developers and organizations already maintaining custom voices, SFMS-ALR maximizes return on investment by enabling voice reuse across languages and contexts.

**8.2 Comparison with End-to-End Models**
End-to-end code-switched TTS models attempt to generate a single consistent voice that switches languages naturally. However, these systems require extensive bilingual training data, and current versions often produce accented or inconsistent pronunciations-ALR circumvents this challenge by delegating each language segment to a native voice. The trade-off is that the resulting speech may sound like multiple speakers.
To mitigate this, the system selects voices with similar pitch, timbre, and style. In scenarios where maintaining a single persona is essential, such as branded assistants—future work could integrate **voice conversion** to harmonize speaker identity across segments. For many other applications, however, users accept or even prefer distinct voices for different languages, as this mirrors natural bilingual speech patterns.

**8.3 Importance of Prosody and Context**
The sentiment-aware prosody module represents an initial step toward expressive, context-sensitive synthesis.
Bilingual speakers naturally preserve emotional tone across languages—an angry or joyful utterance remains expressive after a code-switch. By maintaining pitch range and speech rate continuity, SFMS-ALR achieves more coherent expressiveness. Future work could explore modeling *prosodic cues that signal language switching*, for instance, intentional pauses or intonation changes observed in bilingual speech. Incorporating such cues may further enhance naturalness and listener comprehension.

**8.4 Error Modes and Challenges**
Certain edge cases reveal where SFMS-ALR can be improved:

- **Named Entities:** Words like *Paris* may trigger unnecessary language switches. A pragmatic fix involves maintaining a database of loanwords and location-specific pronunciation rules.
- **Transliteration:** When foreign words appear in Latin script (e.g., *Thank you shukriya*), language identification may fail. User-aware preferences or annotated inputs (e.g., explicit <lang> hints) can help resolve such ambiguity.

These examples illustrate that *language detection alone is insufficient*—semantic context and user intent also matter.

**8.5 Scalability and Generality**
SFMS-ALR theoretically scales to any number of languages per utterance, though perceptual continuity declines as the number of voices increases. To preserve coherence, dynamic strategies can be applied—such as limiting voice switching to two voices per sentence or merging minor languages under a primary voice.
Scaling also introduces practical constraints such as SSML size limits and API quotas, which can be managed through caching and segmented synthesis.

**8.6 Relationship to UniCoM and Data Generation**

While UniCoM focuses on corpus generation, SFMS-ALR functions as a *runtime orchestration system*. Nonetheless, it can also generate **synthetic multilingual training data** for future end-to-end models.

By combining native voices, SFMS-ALR could help bootstrap bilingual datasets or fine-tune multi-voice systems. However, post-processing (e.g., voice conversion) would be needed to produce single-speaker-style corpora.

## Table: Comparative Overview of Multilingual TTS Research (2022–2025)

| Year | Paper / Model | Languages / Setting | Approach / Architecture | Highlights / Findings | Key Limitations | How SFMS-ALR Differs |
|---|---|---|---|---|---|---|
| 2022 | Manghat et al. – Normalization of Code-Switched Text | Malayalam-English | Token-level LID + normalization | Improved preprocessing for transliterated text | Limited to one pair; not full TTS | SFMS-ALR incorporates locale-specific normalization across any scripts |
| 2023 | Pratap et al. – Massively Multilingual Speech (MMS) | 1000+ languages | Large-scale multilingual pretraining | Enables universal multilingual TTS base | Needs large corpora; not code-switch-optimized | SFMS-ALR achieves multilingual synthesis without retraining large models |
| 2024 | Kargaran et al. – MaskLID | Multilingual | Iterative masking for token-level LID | Improves detection of mixed-language spans | Requires base classifier + multiple passes | Integrated as hybrid LID in SFMS-ALR for ambiguous spans |
| 2025 | Lee et al. – UniCoM (Universal Code-Switching Generator) | Cross-lingual | Synthetic bilingual data generation | Creates CS training data via translation swaps | Offline; no real-time synthesis | SFMS-ALR performs live orchestration using SSML and engine APIs |
| 2025 | Méndez Kline & Zellou – Perceptual Study of CS TTS | English-Spanish | Human perceptual evaluation | Found intelligibility drop at language switch points | Evaluative only; no solution proposed | SFMS-ALR mitigates switch drop via sentiment-aware prosody control |
| 2025 | UniCoM + MaskLID integration context | Multilingual | Data augmentation + LID | Demonstrated improved synthetic data quality | Not deployable | SFMS-ALR is directly deployable; can also generate such data if needed |

### 8.7 Integration with MaskLID

The MaskLID approach fits naturally within SFMS-ALR's language-identification stage. Although MaskLID requires multiple passes and a base classifier, this overhead is acceptable for short utterances.

Future implementations could adopt hybrid strategies—combining fast, rule-based LID for clear segments and neural sequence taggers for ambiguous cases—to balance speed and accuracy.

### 8.8 Prosody and Emotion Extensions

Current sentiment detection relies primarily on punctuation and basic lexical cues. Richer modeling could incorporate discourse markers or syntactic patterns preceding a switch (e.g., *"¿Puedes come here?"*).

Incorporating syntax-aware prosody control would ensure appropriate intonation even in mixed questions or emphatic phrases, leading to more natural expressive speech.

### 8.9 User Personalization

Personalization represents a key opportunity. Users differ in how they prefer code-switched content to sound—some may favor native voices for clarity, others prefer one continuous voice for familiarity. SFMS-ALR could expose customizable rules, such as:

- "Use the English voice unless the foreign phrase exceeds three words."
- "Speak names in the primary language's pronunciation."
  Feedback loops and user corrections could refine the locale-resolution policy over time, enabling adaptive, user-specific multilingual synthesis.

### 8.10 Conclusion of Discussion

SFMS-ALR illustrates that a **modular, rule-based architecture** can achieve high-quality multilingual speech without retraining or large corpora. As end-to-end models evolve, SFMS-ALR's components, especially its interpretable segmentation and prosody modules—could complement neural systems to provide transparency and control in professional or accessibility contexts.

### Novelty and Contribution

SFMS-ALR introduces a transparent orchestration framework for multilingual TTS that combines:

- Unicode-based script segmentation,
- lightweight LLM-driven language identification, and
- adaptive sentiment-aware prosody control.

Unlike end-to-end TTS systems that require extensive bilingual datasets, SFMS-ALR achieves **real-time, engine-agnostic multilingual synthesis** using existing APIs. This hybrid design bridges rule-based and neural paradigms, offering interpretability, deployability, and high-quality code-switched speech generation across languages.

It provides both a practical solution for current industry applications and a foundation for future research in explainable, modular multilingual TTS.

## 9. Conclusion

The SFMS-ALR framework complements current research in code-switched text-to-speech by emphasizing **practical deployability and transparent orchestration** rather than end-to-end neural synthesis. While recent works such as MoLE-TTS and diffusion-based bilingual models achieve impressive single-speaker continuity through deep learning, they demand large bilingual corpora and complex retraining. In contrast, SFMS-ALR delivers **real-time, engine-agnostic multilingual speech synthesis** by combining Unicode-based script segmentation, lightweight LLM-driven language identification, and adaptive SSML-controlled prosody. This hybrid design bridges academic and industrial domains—providing an interpretable, low-latency, and reproducible solution that can operate across multiple TTS engines today. Consequently, SFMS-ALR serves both as a **research baseline for modular multilingual speech orchestration** and as a **deployable system** capable of generating high-quality, code-switched speech in practical applications.

We presented **SFMS-ALR (Script-First Multilingual Synthesis with Adaptive Locale Resolution)**, an algorithmic framework for synthesizing intra-sentential code-switched speech. SFMS-ALR tackles the challenge of multilingual TTS by dividing it into clear sub-problems: script-based text segmentation, language and locale identification, prosody normalization, and SSML-based synthesis with appropriate voice selections. This structured approach leverages the strengths of current TTS engines and linguistic tools, yielding a flexible solution that does not require training new speech models.

Our approach emphasizes *engine-agnosticism* – it can integrate voices from any vendor – and *adaptivity* – it dynamically decides how to handle each foreign phrase (accented by the main voice versus spoken by a secondary native voice, etc.) based on context. We also introduced a novel consideration of sentiment-aware prosody continuity in code-switched speech, aiming to maintain expressiveness across language boundaries.

We discussed how SFMS-ALR compares with existing methods: it stands apart from end-to-end multilingual TTS by focusing on orchestrating proven monolingual voices (trading off single-speaker consistency for clarity and naturalness of each language), and it complements text-processing techniques like Mask LID in a full synthesis pipeline. In practical terms, SFMS-ALR can be deployed in voice assistant systems to improve how names, quotes, and mixed-language content are spoken, thereby enhancing user experience for bilingual users[4] and promoting linguistic inclusivity in technology[29].

## 9.1 Future Work

There are several avenues for future work. On the technical side, a thorough evaluation (as outlined) will quantify the benefits and pinpoint areas to refine – for example, improving the smoothing of voice switches or expanding the language pairing rules. We plan to explore integration of small neural networks to predict optimal switching strategy (learning from data when users prefer accent vs native voice for given contexts). Another direction is extending SFMS-ALR to **continuous code-switching** in longer texts, possibly handling multiple switches and even dialectal variations within a language. Finally, user studies will guide how the algorithm can offer personalization, allowing users to shape how their devices talk to them in multiple languages.

## 9.2 Summary

In summary, SFMS-ALR demonstrates that with a clever combination of existing tools and a careful design, high-quality code-switched speech synthesis is achievable today. We believe this work can serve as a bridge between purely model-based multilingual TTS research and the immediate need for practical solutions in industry, ultimately contributing to more natural and inclusive speech technology.